\documentclass[prl,twocolumn]{revtex4}
\usepackage{array,amssymb,graphicx}
\begin{document}

\title{Josephson Effect Without Superconductivity: Realization
in Quantum Hall Bilayers}

\author{Michael M. Fogler and Frank Wilczek}

\affiliation{School of Natural Sciences, Institute for Advanced Study,
Einstein Drive, Princeton, New Jersey 08540}


\raisebox{90pt}[0pt][0pt]{
\mbox{\hspace{4.95in}Preprint IASSNS-HEP-00/54}
}

\vspace{-0.3in}

\begin{abstract}

We show that a quantum Hall bilayer with the total filling $\nu = 1$
should exhibit a dynamical regime similar to the flux-flow in large
Josephson junctions. This analogy may explain a conspicuous peak in the
interlayer tunneling conductance [\prl {\bf 84}, 5808 (2000)]. The
flux-flow is likely to be spatiotemporally chaotic at low bias voltage,
which will manifest itself through broad-band noise. The peak position
can be controlled by an in-plane magnetic field.

\end{abstract}

\maketitle

{\it Motivation\/}.--- In the classic realization of the Josephson
effect, the dynamical variable is the difference between phases of
pairing wave functions in two superconductors. The coupling of this
variable to electromagnetism is dictated by general principles of
gauge invariance. When pairs tunnel from one superconductor to the
other, there is a coherent transport of the charge conjugate to this
phase difference, and the Josephson effect follows. One is led to
exactly the same equations, except with $2e \to e$, when one has a
condensate of excitons -- bound states of particles and holes --
extending across a junction. This is because an exciton, despite its
overall neutrality, does couple to the gauge potential {\it
difference\/} on the two sides of the junction, as a unit charge
$e$. In the exciton condensate the coherent charge transfer is
actually simpler than in a superconductor: it requires a single
tunneling act instead of two~\cite{Comment_on_direct}. Notwithstanding
the hidden connotations, the excitonic condensate can be the ground
state of the junction~\cite{Lozovik_76,Shevchenko_76}. One might be
optimistic that in suitable materials exciton Josephson effects should
be common and could exist at much higher temperatures than
conventional superconductivity, since the fundamental particle-hole
interaction is attractive.

In this Letter we argue that the first observation of such effects has
already occurred albeit in a low-temperature device: a bilayer quantum
Hall system with total filling factor $\nu = 1$. For small interlayer
separations this system spontaneously develops an order parameter
$\Delta \equiv \langle \psi_1^\dagger \psi_2 \rangle \propto e^{i
\phi}$, where $\psi_i^\dagger$ is the electron creation operator in
layer $i$~\cite{Bilayer_review}. When expressed in terms of fermionic
operators $\chi_\uparrow^\dagger \equiv \psi_1^\dagger$, creating an
electron in layer 1, and $\chi_\downarrow^\dagger \equiv \psi_2$,
creating a hole in layer 2, $\Delta$ takes the familiar BCS form
$\langle \chi_\uparrow^\dagger \chi_\downarrow^\dagger \rangle$ and
enables one to identify $\phi$ with the phase of the condensate of bound
electron-hole pairs. We briefly mention that another fruitful analogy
is obtained by treating the layer index as a pseudospin degree of
freedom~\cite{Bilayer_review}. Along that route, nonzero $\Delta$
signifies pseudospin ferromagnetism, while $\phi$ determines the
direction of the in-plane component of the ``magnetization''.

The Josephson effect in the excitonic superfluid was originally
discussed in Ref.~\onlinecite{Shevchenko_76} and more recently in
Ref.~\onlinecite{Lozovik_97}. Its appearance in the quantum Hall
bilayers was also predicted~\cite{Wen_Zee_92,Ezawa_93} but not
immediately accepted~\cite{Anti_Josephson}. Yet a complete formal
analogy to the Josephson junction does hold in the low-energy limit
where $\phi$ obeys a damped-driven sine-Gordon equation of motion,
precisely as the phase variable in a large-area Josephson
junction. The same equation governs many other physical systems, e.g.,
sliding charge-density waves and growing crystals, which implies broad
relevance of the issues discussed below.

The aforementioned likely candidate for the Josephson phenomenon in a
bilayer quantum Hall system is a zero-bias peak in the differential
tunneling conductance $d I / d V$ discovered by Spielman {\it
et al.\/}~\cite{Spielman_00}.
We interpret this peak as an analog of the flux-flow resonance in
a Josephson junction. The discovery of such a
resonance in 1964 by Eck~{\it et al.\/}~\cite{Eck_64} was one of the
first verifications of the Josephson effect in conventional
superconductors. The experiment of Spielman~{\it et al.\/} may well play
a similar role for the quantum Hall bilayers.

Two theoretical papers motivated by this experiment have
appeared~\cite{Stern_xxx,Balents_xxx}. Our independent analysis has some
overlap and some differences with those works. We start with a
perturbative calculation of $I(V)$. One of its predictions is the
dependence of the peak position on the external in-plane magnetic field.
In principle, that can be used to map out the dispersion relation of the
collective mode~\cite{Bilayer_review} analogous to Josephson plasma
oscillations. However, we point out that the perturbation theory
solution is unstable for low damping characteristic of low-temperature
quantum Hall systems, which leads to a spatiotemporally chaotic
flux-flow and large tunneling current fluctuations. Constructing the
theory for such a complicated dynamical regime remains a challenge. For
its numerical study in a discretized formulation
(Frenkel-Kontorova model), see Ref.~\onlinecite{Strunz_98}.

{\it Basic equations\/}.--- The general form of the equation of motion
for $\phi$ can be established from considerations of symmetry and
gauge invariance. To clarify its physical meaning we choose another
approach.  We start with the charge conservation equation $e
\partial_t n + \partial_k j_k + j_{\rm tun} = 0$. The local excess $n$
of the exciton density causes charge imbalance between the layers. We
can express it in terms of the local chemical potential difference
$\mu$ and the capacitance $c$ per unit area: $n = c \mu /e^2$. The
Josephson relation entails $\partial_t \mu = \hbar \partial_t^2
\phi$. The in-plane exciton current $j$ consists of the supercurrent
$j_s = (e \rho_s / \hbar) \partial_x\phi$ and the converted
quasiparticle current $j_{q p} = -\partial_x \mu / e \rho_{x x}$,
where $\rho_s$ is the condensate phase stiffness and $\rho_{x x}$ is
the resistivity of uncondensed quasiparticles. The Josephson tunneling
current is $j_{\rm tun} = (e / \hbar) n_0 \Delta_{\rm sas} \sin(\phi -
Q x_1)$, where $n_0$ is the average electron density per layer,
$\Delta_{\rm sas}$ is the tunneling strength, $Q = e B d / \hbar c$ is
the wavevector imposed by the in-plane component $B$ of the external
magnetic field along the $x_2$-axis. (Its presence is necessitated by
gauge invariance). $d$ is the interlayer separation. Assembling all
the terms, we obtain the equation for $\phi$
(cf.~Ref.~\cite{Bilayer_review}):
\begin{equation}
\left(\partial^2_t - \beta \partial_t \partial_x^2 -
\partial_x^2
\right) \phi + \sin (\phi + \phi_0 - Q x_1) = 0.
\label{sineG}
\end{equation}
Here we expressed distances in units of ``Josephson penetration
length'' $\lambda_J = (\rho_s / n_0 \Delta_{\rm sas})^{1/2}$ and
frequencies in units of ``Josephson plasma frequency'' $\omega_J = v /
\lambda_J$. The velocity parameter $v = (e/\hbar)(\rho_s/c)^{1/2} \sim
0.1 e^2 / \kappa \hbar$ is the one that enters the dispersion
relation~\cite{Wen_Zee_92,Bilayer_review} $\omega^2(q) = \omega^2_J +
v^2 q^2$ of the plasmon and parameter $\beta = (1 / 4 \pi) (\hbar
\omega_J /\rho_s) (\rho_{x x} e^2 / h)$ controls
dissipation~\cite{Comment_on_alpha}. Finally, we included a random
phase shift $\phi_0$ to represent disorder (see below). Using the data
reported in Ref.~\cite{Spielman_00} and theoretical results reviewed
in Ref.~\onlinecite{Bilayer_review} we arrived at rough estimates
$\lambda_J \sim 5\,\mu{\rm m}$, $\omega_J \sim 10^{10}\,s^{-1}$, and
$\beta \sim 0.01$. Boundary conditions for Eq.~(\ref{sineG}) depend on
the sample and measurement geometry.  We consider the case where each
layer is a square with side $L$, the contact to layer 1 is along the
side $x_1 = -L / 2$, $-L / 2 < x_2 < L / 2$, and the contact to layer
2 is along the side $x_1 = L / 2$. This deviates somewhat from the
setup used in Ref.~\onlinecite{Spielman_00} but should be
inconsequencial as long as the bottleneck for the current flow is the
interlayer tunneling not the sheet conductivity. This is indeed the
case experimentally because the inequality $V \gg I \rho_{xx}$ is
satisfied ($V$ is the voltage difference between the contacts). Under
such conditions it is also permissible to choose $\phi_L = \phi_R = V
t$, and use
\begin{equation}
  I(t) = \int d x_2 [\partial_1 \phi_R(t) - \partial_1 \phi_L(t)].
\label{bc_grad_phi}
\end{equation}
to calculate the tunneling current. Here $\phi_{L, R} \equiv \phi(x_1 =
\pm L / 2, x_2)$ and our units of voltage and electric current are $V_0
= \hbar \omega_J / e$ and $I_0 = e \rho_s / \hbar$, respectively.

Before we proceed to the calculations let us explain the origin the the
term ``flux-flow.'' Let us consider the case $\phi_0 = \partial_t \phi =
V = 0$ and focus on the limit $Q \gg 1$ where the stable (ground-state)
solution of Eq.~(\ref{sineG}) corresponds to an almost uniform phase
distribution $|\partial_1\phi| \ll 1$. There is an equivalent
alternative formulation in terms of a shifted phase $\theta \equiv \phi -
Q x_1$ in which $Q$ disappears from the argument of the sine but
re-emerges in the boundary conditions for $\partial_1 \theta = \partial_1
\phi - Q$. In this formulation the ground-state $\theta$ varies rapidly
and almost linearly in space, which can be described as the $2 \pi /
Q$-periodic lattice of $2\pi$-solitons. A nonequilibrium state for $V
\neq 0$ where the phase increases uniformly with the rate $V$ can then
be visualized as a uniform sliding of the soliton lattice --- whence the
term flux-flow.

{\it Perturbation theory\/}.--- The perturbative solution of
Eq.~(\ref{sineG}) for $\phi_0 = 0$ is readily done in terms
of the Green's function
\begin{equation}
  G(\omega, k) = [\omega^2 + i \alpha(k) \omega - k^2]^{-1},
\quad \alpha(k) \equiv \beta k^2.
\label{g}
\end{equation}
The dc current $\bar{I}$ is then obtained by averaging
Eq.~(\ref{bc_grad_phi}) over time. The full expression is somewhat
cumbersome but for $L$ in the range $1 \ll L \ll \bar{I}^{-1}$ it
reduces to~\cite{Eck_64}
\begin{equation}
\bar{I} = \frac{L^2}{2}
\frac{\alpha(Q) V}{(V^2 - Q^2)^2 + \alpha^2(Q) V^2}.
\label{I_ff}
\end{equation}
We estimate $L \sim 50$ and $\bar{I} \sim 10^{-6}$ for the conditions
of Ref.~\onlinecite{Spielman_00}, so $L$ is in the required
range. From Eq.~(\ref{I_ff}) we see that the resonance arises when the
velocity $V / Q$ of the soliton train matches the plasmon velocity $v$
($v = 1$ in the adopted dimensionless units). Under this
condition the power dissipation in the system reaches a maximum. More
generally, the resonance condition is $e V = \hbar \omega_0(Q)$ where
$\omega_0(Q)$ is the $\Delta_{\rm sas} \to 0$ limit of the plasmon
dispersion relation (in which the plasmon becomes the Goldstone
mode). The last conclusion was reached independently in
Refs.~\onlinecite{Stern_xxx} and \onlinecite{Balents_xxx}.

Let us turn to the disordered case, $\phi_0 \neq 0$. From
Eq.~(\ref{sineG}) and the identity $\sin z = {\rm Im}\,e^{i z}$, one can
see that the random phase factor $e^{i \phi_0}$ plays the same role for
the disordered system as the periodic phase factor $e^{i Q x_1}$ for the
clean system. The Fourier expansion of $e^{i \phi_0}$ can be thought
of as a set of ``diffraction gratings'' with different wavevectors $k$,
resonating whenever $V = k$. As a result, the resonance is broadened in
proportion to the breadth of the Fourier spectrum of the
following correlator:
\begin{equation}
\sigma({\bf x}) = \langle e^{-i \phi_0({\bf x})} e^{i\phi_0(0)}
\rangle,
\label{sigma}
\end{equation}
and the appropriate generalization of Eq.~(\ref{I_ff}) is
\begin{equation}
 \bar{I} = -\frac{L^2}{2}\, {\rm Im} \int \frac{d^2 k}{(2 \pi)^2}
 \tilde\sigma({\bf k} - {\bf Q}) G(V, k).
\label{I_ff_disorder}
\end{equation}
To evaluate the integral we assume that the random phase field $\phi_0$
is mainly due to static randomly positioned and randomly oriented bound vortex
pairs in the exciton condensate. In the context of the pseudospin
ferromagnet analogy mentioned earlier~\cite{Bilayer_review}, such
vortices were recognized as merons, the topological defects of the
$O(3)$ nonlinear $\sigma$-model, and a remarkable fact was established:
each meron possesses an overall electric charge $e / 2$ concentrated
near its core. The competition of the Coulomb repulsion between the like
electric charges and attraction between oppositely charged vortices
selects the optimal size $a$ of meron pairs; $a$ remains finite below
the roughening transition~\cite{Chui_76} temperature $T_r = 8 \pi \rho_s
/ k_B$. From our point of view, in the context of the Josephson junction
analogy, the meron pairs correspond to misaligned Abrikosov vortices
trapped in the junction. The $I$-$V$ characteristic of such junctions
was investigated by Fistul and Giuliani~\cite{Fistul_97}. These authors
also attempted to go beyond the perturbation theory but that part of
their calculation is suspect for reasons that will become clear shortly.

Within the chosen model of disorder we obtain
\begin{eqnarray}
\tilde\sigma(k) &=& A a^2 (k a)^{\gamma - 2},\quad
\gamma \equiv \pi n_m a^2,
\label{sigma_k}\\
A &\simeq& 2 \pi \gamma,\quad\quad\quad\quad\:\:
\gamma \ll 1,
\label{A_small_gamma}\\
A &\simeq& 2 \pi \ln (k a)^{-1},\quad \gamma = 2,
\label{A_gamma_2}
\end{eqnarray}
where $n_m$ is average density of the meron pairs.
Substituting the above formulas into Eq.~(\ref{I_ff_disorder}) yields
for $Q = 0$ and $V_* \ll V \ll \beta^{-1}$:
\begin{equation}
 \bar{I} = I_* (V_* / V)^{2 - \gamma},\quad
     I_* = L^2 \alpha(V_*) V_*.
\label{I_ff_1}
\end{equation}
Here $V_* = [A(V_*) a^\gamma / 8 \beta]^{1 / (5 - \gamma)}$ is
determined from the condition that the variance of $\phi$, which is
given by
\begin{equation}
 \langle \phi^2 \rangle  = \frac{1}{2} \int \frac{d^2 k}{(2 \pi)^2}
 \tilde\sigma({\bf k} - {\bf Q}) |G(V, k)|^2,
\label{phi_rms}
\end{equation}
becomes of the order of one. At $V < V_*$ the premise of the
perturbation theory ($\phi \ll 1$) is violated and
formula~(\ref{I_ff_1}) is invalid. However, $V = V_*$ is not the true
boundary of the perturbative domain. Below we will show that the
perturbation theory actually breaks down in a wider range of voltages,
$V < V_{**}$, where $V_{**} \gg V_*$ for small $\beta$.

\begin{figure}
\includegraphics[width=1.8in,bb=129 207 429 434]{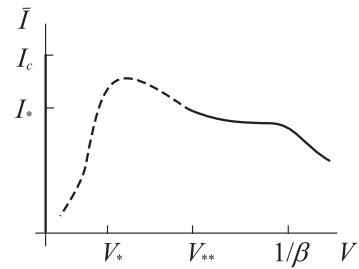}
\vspace{0.1in}
\setlength{\columnwidth}{3.2in}
\caption{
$I$-$V$ characteristic given by the perturbation theory with $\gamma \approx 2$
(solid lines). The dashed line is a conjecture.
\label{Fig_I_V}
}
\end{figure}

In the following we focus on the case $\gamma \approx 2$ (not an
unreasonable value~\cite{Comment_on_gamma}) where the Fourier spectrum
of $e^{i \phi_0}$ is extremely broad, see Eq.~(\ref{sigma_k}). The
flux-flow resonance is equally broad so that it shows up as a plateau
in $\bar{I}(V)$, see Fig.~\ref{Fig_I_V}. With the chosen value of
$\gamma$ we get the expression (physical units temporarily restored)
\begin{equation}
   I_* \sim (e \rho_s / \hbar) (L^2 a^2 / \lambda_J^4)
\label{I_*}
\end{equation}
for the height of the plateau. Similar expressions were also derived
in Refs.~\onlinecite{Stern_xxx} and \onlinecite{Balents_xxx}.

Even though the response at low $V$ is beyond the reach of the
perturbation theory, the ``Josephson critical current,'' which is a
zero-bias parameter, can nevertheless be calculated. The
collective-pinning theory of Vinokur and Koshelev~\cite{Vinokur_90}
yields (for $\gamma = 2$)
\begin{equation}
 I_c = 2 (e \rho_s / \hbar) (L^2 a^2 / \lambda_J^4)
 \ln^2 (\lambda_J / a).
\label{I_c}
\end{equation}
Reminiscent of the inequality between the coefficients of static and
kinetic friction, $I_c$ exceeds $I_*$. Note that $I_c$ is essentially a
static quantity. It need not coincide with the $V \to 0$ limit
of the dynamical response formulas akin to Eq.~(\ref{I_ff_disorder}).
The overall dependence of $\bar{I}$ on $V$ is likely to appear the way
shown in Fig.~\ref{Fig_I_V}.

%
%

Choosing $\gamma \approx 2$ enables us to reproduce the plateau feature
in the experimental data~\cite{Spielman_00} (the peak in $d I / d V$
corresponds to a plateau in $\bar{I}$ {\it vs\/} $V$), but the
theoretical estimate of $I_*$ is off by two orders of magnitude.
However, this estimate is highly unreliable in view of large
uncertainties in the parameters of the model and of their strong
renormalization by thermal fluctuations, known from the theory of the
roughening transition~\cite{Chui_76}.

\begin{figure}
\includegraphics[width=2.1in,bb=136 359 452 530]{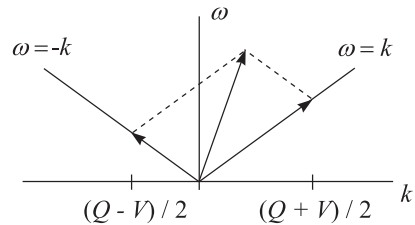}
\vspace{0.1in}
\setlength{\columnwidth}{3.2in}
\caption{Kinematic diagram for the parametric resonance.
\label{Fig_kinematics}
}
\end{figure}

{\it Instability of a tachyonic flux-flow\/}.--- It is known, although
not widely appreciated, that the perturbation theory solution for the case
$\phi_0 =
0$ -- in the form of a moving soliton lattice -- can become
unstable when the lattice velocity exceeds the
velocity of Josephson plasmon, $V / Q > 1$. The instability can be
understood as a parametric resonance driven by the external source of
frequency $V$ and wavevector $Q$. Indeed, if $V > Q$, then kinematics
allows a simultaneous excitation of two counter-propagating plasma
waves, with wavevectors $k_\pm = (Q \pm V) / 2$, see
Fig.~\ref{Fig_kinematics}.

%
%

For large $Q$ the instability appears
when~(cf.~Ref.~\onlinecite{Malomed_91})
\begin{equation}
     G^{-1}(k_+, k_+) G^{-1}(-k_-, k_-) < 1 / 4,\quad V > Q,
\label{instability_criterion}
\end{equation}
and so the perturbation theory solution on the ``tachyonic'' side $V >
Q$ is stable only if $V > V_{**} = (4 / \beta)^{1/3}$ ($Q \ll V_{**}$ is
assumed). This condition is much more restrictive than the naive $V - Q
\gg Q^{-1}$ derived from the criterion $|\phi| \ll 1$.

\begin{figure}
\includegraphics[width=1.9in,bb=205 227 448 596]{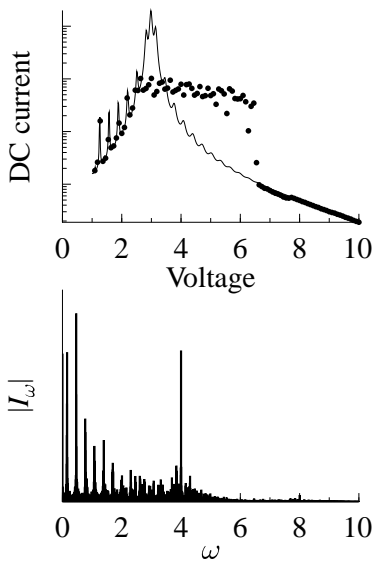}
\vspace{0.1in}
\setlength{\columnwidth}{3.2in}
\caption{
Top: $I$-$V$ characteristic. Dots are from numerics, solid line is
from a full perturbation theory formula for a finite-length
system. It contains both the main resonance of
Eq.~(\protect\ref{I_ff}) and smaller ``Fiske resonances.''
The scatter of the datapoints at $V > 2.5$ is due to statistical
noise. The parameters of simulation are $L = 20$, $Q =
3$, $\beta = 0.01$, and $\phi_0 = 0$.
Bottom: An example of the frequency spectrum of $I(t)$
in the above simulation. Note the narrow peak at $\omega$ equal
to the applied voltage $V = 4$ and
the strong broad-band noise at smaller $\omega$'s.
\label{Fig_chaos}
}
\end{figure}

Very close to the threshold $V_{**}$, the parametric resonance
described above produces a small modulation of the uniform soliton
train~\cite{Malomed_91}, but as $V$ moves closer to $Q$ other instable
modes proliferate and the system dynamics quickly enters the regime of
spatiotemporal chaos. In Fig.~\ref{Fig_chaos} we show the comparison
between the results of the naive perturbation theory and of our
numerical simulations of a one-dimensional system. As expected, they
agree when either $V \gtrsim V_{**}$ or $Q - V \gtrsim Q^{-1}$, but at
intermediate $V$'s they differ substantially and a strong broad-band
noise in $I$ sets in.

%
%

Let us now show that a similar instability must occur in the
disordered case as well. To estimate the corresponding $V_{**}$ we
approximate the unstable modes by wavepackets of plasma waves with a
Gaussian spread of wavevectors, $\psi_\pm({\bf k}) = (2 \pi \Delta
k^2)^{-1} \exp [-|{\bf k} - {\bf k}_\pm|^2 / 2 (\Delta k)^2]$. If
$\Delta k$ is smaller than $\alpha(k_\pm)$, then $G(k, {\bf k})$ is
approximately the same for dominant ${\bf k}$'s within the
wavepackets, and so the left-hand side of the instability
criterion~(\ref{instability_criterion}) need not be modified. The
right-hand side however is changed from $1 / 4$ to $(1 / 2 \pi)
\tilde\sigma(V - Q) (\Delta k)^2$ and the threshold voltage becomes $
V_{**} \sim \beta^{-1/2} \tilde\sigma^{-1/8}$. Thus, for small $\beta$
the lowest voltage at which the perturbation theory is expected to
apply is $V_{**} \gg V_*$. Our conjecture on the behavior of $\bar{I}$
in the interval $V_* < V < V_{**}$ (depicted in Fig.~\ref{Fig_I_V}) is
based on a notion that chaos leads to a higher effective dissipation,
see Fig.~\ref{Fig_chaos} (top) and Refs.~\cite{Strunz_98} and
\cite{Malomed_91}. The suppression of $\bar{I}$ below $V_*$ is an
educated guess motivated by the work of Fistul and
Giuliani~\cite{Fistul_97}.

The probable observation of Josephson effects without bulk
superconductivity in a bilayer quantum Hall ferromagnet paves the way
for exploring a vast variety of other Josephson phenomena in this
system. It should also inspire attempts to realize them in other
systems. The complexity and appeal of the emerging theoretical issues
(not fully appreciated in the current literature), as well as their
recurrence in other contexts warrant further study. From this
perspective the lack of quantitative agreement between the
perturbation theory and experiment is stimulating.

M.~M.~F.~ is indebted to Jim~Eisenstein for illuminating
discussions and to Aspen Center for Physics for hospitality. This
research is supported by US DOE Grant No.~DE-FG02-90ER40542.


\begin{thebibliography}{19}
\bibitem{Comment_on_alpha} Another possible term in Eq.~(\ref{sineG})
$\alpha \partial_t \phi$ (due to incoherent interlayer tunneling
of quasiparticles) is more relevant by power counting but is
suppressed by the Coulomb blocking of
tunneling~\cite{Bilayer_review}, i.e., $\alpha \lll 1$.
\end{thebibliography}
\end{document}